\documentstyle{article}
\begin{document}
\hoffset -1.5cm

\title{\Large\bf
A [SU(6)]$^4$ FLAVOR MODEL WITHOUT MIRROR FERMIONS}
\author{\large William A. Ponce$^{1,2}$ and Arnulfo Zepeda$^2$\\
\normalsize 1-Departamento de F\'\i sica, Universidad de Antioquia \\
\normalsize A.A. 1226, Medell\'\i n, Colombia.\\
\normalsize 2-Departamento de F\'\i sica,
Centro de Investigaci\'on y de Estudios Avanzados del IPN.\\
\normalsize Apartado Postal 14-740, 07000 M\'exico D.F., M\'exico.\\}
\date{November 13, 1993}

\maketitle
%VERSION DE NOVIEMBRE 13 PARA EL PHYS. REV., ENVIADA AL Z PHYS. C EL 4 DE
%FEBRERO POR DHL.
\vspace{2.cm}

\renewcommand{\baselinestretch}{1.2}
\setlength{\hsize}{15cm}
\setlength{\vsize}{10in}

\newcommand{\sm}{SU(3)$_C\otimes$SU(2)$_L\otimes$U(1)$_Y$}
\newcommand{\sem}{SU(3)$_C\otimes$U(1)$_{EM}$}

\large

\begin{center}
{\bf ABSTRACT}
\end{center}

\parbox{15cm}%
{We introduce a three family extension of the
Pati-Salam model which is anomaly-free and contains in a single irreducible
representation the known quarks and leptons without mirror fermions.
Assuming that the breaking of the symmetry admits the implementation of the
survival
hypothesis, we calculate the mass scales using the
renormalization group equation. Finally we show that the proton remains
perturbatively stable.}

\pagebreak

\section{INTRODUCTION}
The renormalizability of the Pati-Salam\cite{ps} (PS) model for
unification of flavors and forces rest on the existence of conjugate or
mirror partners of ordinary fermions. Mirror fermions are fermions with
quantum numbers with respect to the Standard Model (SM) gauge group \sm
identical to those of the known quarks and leptons, except that they
have opposite handedness from ordinary fermions. Their existence vitiate
the survival hypothesis\cite{sh} according to which chiral fermions that
can pair off while respecting a symmetry will do so, acquiring
masses grater than or equal to the mass scale of the respected symmetry.

These remarks are illustrated in one of
the PS type models. The gauge group for the
three-family extension of the PS model is\cite{elias}

\begin{center}
G$^\prime\equiv$
SU(6)$_L\otimes$SU(6)$_R\otimes$SU(6)$_{CL}\otimes$SU(6)$_{CR}\times$Z$_4$,
\end{center}

\noindent
where $\otimes$ indicates a direct product, $\times$ a semidirect one,
and Z$_4\equiv$(1,P,P$^2$,P$^3$) is the four-element cyclic group acting
upon [SU(6)]$^4$ such that if (A,B,C,D) is a representation of
[SU(6)]$^4$ with A a representation of the first factor, B of the second,
C of the third, and D of the fourth, then P(A,B,C,D)=(B,C,D,A) and then
Z$_4$(A,B,C,D)$\equiv $(A,B,C,D) $\oplus$ (B,C,D,A) $\oplus$ (C,D,A,B)
$\oplus$ (D,A,B,C). The charge operator in G$^\prime$ is defined
as\cite{elias}

\begin{equation}
Q_{EM}=T_{ZL}+T_{ZR}+Y_{(B-L)_L}+Y_{(B-L)_R}.
\end{equation}

\noindent
The irreducible representation (irrep) of G$^\prime$ which contains
the known fermions is

\begin{center}
$\psi^\prime(144)\equiv {\rm Z}_4\psi^\prime(\bar{6},1,6,1)=
\psi^\prime(\bar{6},1,6,1)\oplus\psi^\prime(1,6,1,\bar{6})
\oplus\psi^\prime(6,1,\bar{6},1)\oplus\psi^\prime(1,\bar{6},1,6)$,
\end{center}

\noindent
where $\psi^\prime(\bar{6},1,6,1)$ includes the known left-handed
weak doublets while $\psi^\prime(1,6,1,\bar{6})$ includes the known
right-handed weak singlets of the three families.
$\psi^\prime(6,1,\bar{6},1)$  and $\psi^\prime(1,\bar{6},1,6)$
are the mirror fermions of $\psi^\prime(\bar{6},1,6,1)$ and
$\psi^\prime(1,6,1,\bar{6})$ respectively.
With this particle content G$^\prime$ is free of anomalies because
the mirror multiplets cancel the anomalies introduced by the multiplets
which contain the known fermions.
The model defined  with G$^\prime$ and $\psi^\prime (144)$ does not have a
symmetry that would forbid mass terms of the
form $\psi^\prime(\bar{6},1,6,1)\psi^\prime(6,1,\bar{6},1)+
\psi^\prime(1,6,1,\bar{6})\psi^\prime(1,\bar{6},1,6)$ at the
G$^\prime$ scale. The aim of the present work is to introduce a
variation of this PS model.

A change in the definition of the permutation operator P induces a change
in Z$_4$ and therefore in the definition of G$^\prime$.
The new group, G, can be written also with the same Z$_4$ as in G$^\prime$
but interchanging the order of the factor groups. In this notation we will
consider the gauge group

\begin{center}
G=SU(6)$_L\otimes$SU(6)$_R\otimes$SU(6)$_{CR}\otimes$SU(6)$_{CL}\times$Z$_4$
\end{center}
\noindent
with
\begin{center}
$\psi(144)=$Z$_4\psi(\bar{6},1,1,6)=\psi(\bar{6},1,1,6)\oplus
\psi(1,1,6,\bar{6})\oplus\psi(1,6,\bar{6},1)\oplus\psi(6,\bar{6},1,1)$.
\end{center}
This gauge structure is also free of anomalies but has a different
particle content. Indeed, the ordinary fermions in $\psi(144)$
are included now in $\psi(\bar{6},1,1,6)\oplus\psi(1,6,\bar{6},1)$, but
$\psi(6,\bar{6},1,1)\oplus\psi(1,1,6,\bar{6})$ does not contain the mirror
fermions of the ordinary fermion fields. To see this let us write the
quantum numbers for $\psi(144)$ with respect to the SM group
[our notation designates transformation behavior under
(SU(3)$_C$, SU(2)$_L$, U(1)$_Y$)]:

\noindent
$\psi(\bar{6},1,1,6)\sim 3(3,2,1/3)\oplus 6(1,2,-1)\oplus3(1,2,1)$\\
$\psi(1,6,\bar{6},1)\sim 3(\bar{3},1,-4/3)\oplus 3(\bar{3},1,2/3)\oplus
6(1,1,2)\oplus 9(1,1,0)\oplus 3(1,1,-2)$\\
$\psi(6,\bar{6},1,1)\sim 9(1,2,1)\oplus 9(1,2,-1)$\\
$\psi(1,1,6,\bar{6})\sim (8+1,1,0)\oplus 2(3,1,4/3)\oplus 2(\bar{3},1,-4/3)
\oplus (3,1,-2/3)\oplus (\bar{3},1,2/3)\oplus 5(1,1,0)\oplus 2(1,1,2)
\oplus 2(1,1,-2),$\\
where the ordinary left-handed quarks correspond to 3(3,2,1/3) in
$\psi(\bar{6},1,1,6)$, the ordinary right-handed quarks correspond to
3($\bar{3},1,-4/3)\oplus 3(\bar{3},1,2/3)$ in $\psi(1,6,\bar{6},1)$,
the known left-handed leptons are in three of the six (1,2,$-1$) of
$\psi(\bar{6},1,1,6)$, and the known right-handed charged leptons are in
three of the six (1,1,2) of $\psi(1,6,\bar{6},1)$. The
exotic leptons in $\psi(\bar{6},1,1,6)$ belong to the vectorlike
representation $3(1,2,-1)\oplus 3(1,2,1)$ (vectorlike with respect to
the SM quantum numbers) and the exotic leptons in $\psi(1,6,\bar{6},1)$
belong to the vectorlike representation $3(1,1,2)\oplus 3(1,1,-2)\oplus
9(1,1,0)$, where three lineal combinations of the nine states with
quantum numbers (1,1,0) can be identified as the
right-handed neutrinos.

Notice that the G symmetry and the representation content of
$\psi(144)$ forbid mass terms for fermion fields
at the unification scale, but according to
the survival hypothesis the vectorlike substructures pointed in the
former and in the next paragraphs should get masses one scale above
M$_Z$, the known weak interactions mass scale.

$\psi(6,\bar{6},1,1)$ is formed by 36 exotic Weyl leptons, 9 with
positive electric charges, 9 with negative (the charge conjugates to
the positive ones), and 18 are neutrals; all together constitute a
vectorlike representation. Also all the particles in
$\psi(1,1,6,\bar{6})$ form a vectorlike representation, where
$5(1,1,0)\oplus 2(1,1,2)\oplus 2(1,1,-2)$ stand for nine exotic
leptons (electric charges 0, $\pm 1$),
$2(3,1,4/3)\oplus 2(\bar{3},1,-4/3)$ refers to two exotic UP
type quarks (electric charge 2/3), $(3,1,-2/3)\oplus (\bar{3},1,2/3)$
refers to one exotic DOWN type quark (electric charge $-1/3$), and the
nine states in (8+1,1,0) are the most exotic, electrically neutral
fermion fields in the model, whose origin and meaning is discussed anon.

The model described by [G, $\psi(144)$] (or either by [G$^\prime$,
$\psi^\prime (144)$]) unifies the three family SM gauge group, and it
unifies also the more general three family chiral
color extension of the SM, which has the gauge structure\cite{gf}

\begin{center}
R$\equiv$SU(3)$_{CR}\otimes$SU(3)$_{CL}\otimes$SU(2)$_L\otimes$U(1)$_Y$,
\end{center}

\noindent
where the unbroken color group SU(3)$_C$ of the SM is identified with the
diagonal subgroup of SU(3)$_{CR}\otimes$SU(3)$_{CL}$. The model described
by [G, $\psi(144)$] is an alternative to the PS
model for three families and it is a unified theory of a new chiral model
with special features, different from the models presented in
Refs.\cite{gf}. The nine states (8+1,1,0) in $\psi(1,1,6,\bar{6})$
are related to the so-called dichromatic fermion multiplets,
belonging to the $(3,\bar{3})$ representation of the
SU(3)$_{CR}\otimes$SU(3)$_{CL}$ subgroup of R. Then,
according to the nomenclature introduced in Ref.\cite{gf},
(8+1,1,0) is formed
by the {\it ``queight"} (8,1,0) and the color neutral {\it ``quone"}
(1,1,0).

Another feature of the model described by [G, $\psi(144)$] is
that it is the chiral color extension of the vector-color-like
model described by
G$^V\equiv $ SU(6)$_L\otimes$SU(6)$_C\otimes$SU(6)$_R\times$Z$_3$ and
$\psi^V(108)$=Z$_3\psi^V(\bar{6},6,1)
=\psi^V(\bar{6},6,1)\oplus\psi^V(6,1,\bar{6})\oplus\psi^V(1,\bar{6},6)$.
This vector like model was sketched for
the first time in Ref.\cite{mex} and studied in detail in Refs.\cite{ponce}.
SU(6)$_C$ in G$^V$ is the diagonal subgroup of
SU(6)$_{CR}\otimes$SU(6)$_{CL}$ in G, and the particle content of
the two models is almost the same in the following sense:
$\psi^V(\bar{6},6,1)=\psi(\bar{6},1,1,6)$,
$\psi^V(1,\bar{6},6)=\psi(1,6,\bar{6},1)$, and
$\psi^V(6,1,\bar{6})=\psi(6,\bar{6},1,1)$. Hence, several techniques used
and some results obtained in the study\cite{ponce} of the structure
[G$^V$, $\psi^V(108)$] can be translated to the
study of [G, $\psi(144)$].

\section{SYMMETRY BREAKING}
Let us break G down to \sem by the introduction of appropriate elementary
Higgs fields which trigger the spontaneous breaking of the symmetry and
at the same time produce masses for the fermion fields in $\psi(144)$, in such
a way that the survival hypothesis\cite{sh} holds at each mass scale.

First let us consider the two mass scale symmetry breaking pattern

\begin{center}
G$\stackrel{M}{\longrightarrow}$\sm$\stackrel{M_Z}{\longrightarrow}$\sem,
\end{center}

\noindent
with M$>>$M$_Z$. The running coupling constants of the SM satisfy
the one loop Renormalization Group Equations (RGEs)

\begin{equation}
\alpha^{-1}_i(M_Z)=\alpha^{-1}_i(M)-b_i{\rm ln}(M/M_Z),
\label{eq1}
\end{equation}

\noindent
where $\alpha_i=g_i^2/4\pi, i=1,2,3$ refers to U(1)$_Y$, SU(2)$_L$ and
SU(3)$_C$ respectively, and

\begin{equation}
b_i=\{\frac{11}{3}C_i(vectors)
-\frac{2}{3}C_i(Weyl-fermions)-\frac{1}{6}C_i(scalars)\}/4\pi,
\label{eq2}
\end{equation}

\noindent
where $C_i(...)$ is the index of the representation to which the
(...) particles are assigned (for a complex scalar field the values
of $C_i(scalars)$ should be doubled).

With the normalization of the generators in G such that
$\alpha_1(M)=\alpha_2(M)=2\alpha_3(M)=\alpha_{CL}(M)=\alpha_{CR}(M)
\equiv\alpha$ (where $\alpha_{CL(CR)}=g^2_{CL(CR)}/4\pi$ refers to
the gauge coupling constant for SU(3)$_{CL(CR)}$ in R), the relationship

\begin{equation}
\alpha_{EM}=\frac{1}{3}\alpha_2{\rm sin}^2\theta_W=\frac{3}{19}
\alpha_1{\rm cos}^2\theta_W,
\label{eq3}
\end{equation}

\noindent
where $\theta_W$ is the weak mixing angle, is valid at all energy scales.
This last equation implies also that at all energies

\begin{equation}
3\alpha^{-1}_{EM}=19\alpha^{-1}_1+\alpha^{-1}_2.
\label{eq4}
\end{equation}

\noindent
{}From the former equations we get

\begin{equation}
\frac{3}{28}\alpha^{-1}_{EM}(M_Z)=\frac{\alpha^{-1}_3(M_Z)}{2}+
(\frac{b_3}{2}-\frac{9b_2}{28}-\frac{19b_1}{28}){\rm ln}(M/M_Z)
\label{eq5}
\end{equation}

\noindent
and

\begin{equation}
{\rm sin}^2\theta_W(M_Z)=3\alpha_{EM}(M_Z)
\{\frac{\alpha_3^{-1}(M_Z)}{2}+(\frac{b_3}{2}-b_2){\rm ln}(M/M_Z)\}.
\label{eq6}
\end{equation}

\noindent
After decoupling the vector-like representations in $\psi(144)$
according to the Appelquist-
Carazzone theorem\cite{apel} we get: $2\pi b_3=[7-C_3(s)/12]$,
$2\pi b_2=[10/9-C_2(s)/36]$ and $2\pi b_1=-[20/19+C_1(s)/76]$,
where $C_i(s),i=1,2,3$ are the indices for the Higgs fields contributing to
$b_i$. Now, the set of Higgs fields needed to break G down to
\sm and to give at the same time
masses to the vectorlike fermions in $\psi(144)$ contribute negligible to
$C_i(s)$, because in the effective theory their contribution is highly
suppressed by powers of M$_Z$/M. So the only Higgs fields in existence
below M are those which break \sm down to \sem.
The simplest set of Higgs fields and Vacuum Expectation Values (VEVs)
which do the last breaking and at the same time give rise to mass terms
for the known fermion fields is
$\phi_1(72)=\phi_1(6,1,\bar{6},1)\oplus\phi_1(1,\bar{6},1,6)
=\phi^a_\Delta\oplus\phi_A^\alpha$ (where a,b,...;A,B,...$\alpha,\beta,...
\Delta,\Omega,...$ = 1,...6 label SU(6)$_{L}$, SU(6)$_{R}$, SU(6)$_{CL}$ and
SU(6)$_{CR}$ tensor indices respectively), with VEVs
$\langle\phi_1(72)\rangle\neq 0$ in the directions
(a,$\Delta$) = (2,4) = (4,4) = (6,4) = (1,5) = (3,5) = (5,5) = (2,6) = (4,6)
= (6,6), and ($\alpha$,A) = (4,2) = (4,4) = (4,6) = (5,1) = (5,3) = (5,5)
= (6,2) = (6,4) = (6,6). It can be seen that this set is inconsistent
with the known quark mass spectrum because it generates see-saw
masses for the t and b quarks of the same order of magnitude and
proportional to M$_Z^2$/M. The alternative is to look for a set of Higgs
fields and VEVs which breaks the symmetry and
generates at the same time, what is called in Refs.\cite{ponce} the
{\it ``modified horizontal survival hypothesis"}, according to which the t
quark gets a mass of order M$_Z$ via a flavor democratic mass matrix,
with the hope that different see-saw mechanisms\cite{seesaw}
and radiative corrections\cite{ma} reproduce the hierarchy of masses and
mixing angles for quarks and leptons. This scenario can be achieved
by using, besides $\phi_1(72)$, other set of Higgs fields
$\phi_2(1296)=\phi_2(6,\bar{6},6,\bar{6})\equiv\phi^{a,\Delta}_{A,\alpha}$
with VEVs such that (a,A) = (2,2) = (2,4) = (2,6) = (4,2)
= (4,4) = (4,6) = (6,2) = (6,4) = (6,6),
and ($\Delta ,\alpha)$ = (1,1) = (2,2) = (3,3) = (4,4) = (5,5) = (6,6).
($\phi_2$ with the VEVs as stated here not only does the job as
desired but it also breaks SU(3)$_{CR}\otimes$SU(3)$_{CL}$ in R  down
to SU(3)$_C$.)

But how many of the 72 Higgs fields in $\phi_1$ and
of the 1296 Higgs fields in $\phi_2$ contribute to $C_i(s)$? Let us
work with two hypothesis:\\
{\it Hypothesis i}. All the Higgs fields in $\phi_1(72)\oplus\phi_2(1296)$
contribute to $C_i(s)$. It is easy to show that in this case the Higgs field
contribution to Eqs.(\ref{eq5}) and (\ref{eq6}) cancels out. (That the
contribution of $\phi_1$ (and also of $\phi_2$ separately) cancels out in
Eqs.(\ref{eq5}) and (\ref{eq6}) can also be seen from general principles.)
Substituting the experimental values\cite{amal} sin$^2\theta_W(M_Z)
=0.2341\pm 0.0025$, $\alpha^{-1}_{EM}(M_Z)=127.6\pm 0.2$
and $\alpha_3(M_Z)=0.122\pm 0.005$, we get from Eq.(\ref{eq5})
ln(M/M$_Z$)=15.59$\pm 0.31$ while from Eq.(\ref{eq6})
ln(M/M$_Z$)=15.45$\pm 0.76$. The compatibility of these results
with each other allows us to obtain M=5.5$\times 10^6$ GeVs and
to claim that with this hypothesis and with G breaking down to \sem
with the SM gauge group as the only intermediate gauge structure,
the four coupling
constants meet together at a single point M. Unfortunately for
this scheme any new physics is at the mass scale M$\sim 10^6$ GeVs.

{\it Hypothesis ii}. Only the Higgs fields which develop VEVs contribute to
$C_i(s)$ (hypothesis known in the literature as the {\it ``extended
survival hypothesis"}\cite{esh}). Under this assumption we get from
Eq.(\ref{eq5}) ln(M/M$_Z)=12.48\pm 0.25$ while from Eq.(\ref{eq6})
ln(M/M$_Z)=10.89\pm 0.54$ which are inconsistent solutions.

The other symmetry breaking pattern with only one intermediate mass scale,
consistent with present experiments\cite{gf} is

\begin{center}
G$\stackrel{M}{\longrightarrow}$R=SU(3)$_{CR}\otimes$
SU(3)$_{CL}\otimes$SU(2)$_L\otimes$U(1)$_Y
\stackrel{M_Z}{\longrightarrow}$\sem,
\end{center}

\noindent
where again M$>>$M$_Z$.
To study this case let us write the quantum numbers for $\psi(144)$ with
respect to R, [now our notation designates transformation behavior under
(SU(3)$_{CR}$, SU(3)$_{CL}$, SU(2)$_L$, U(1)$_Y$)]\\
$\psi(\bar{6},1,1,6)\sim 3(1,3,2,1/3)\oplus 6(1,1,2,-1)\oplus 3(1,1,2,1)$\\
$\psi(1,6,\bar{6},1)\sim 3(\bar{3},1,1,-4/3)\oplus 3(\bar{3},1,1,2/3)\oplus
6(1,1,1,2)\oplus 3(1,1,1,-2)\oplus 9(1,1,1,0)$\\
$\psi(6,\bar{6},1,1)\sim 9(2,1,1,1)\oplus 9(2,1,1,-1)$\\
$\psi(1,1,6,\bar{6})\sim (3,\bar{3},1,0)\oplus 2(1,\bar{3},1,-4/3)\oplus
(1,\bar{3},1,2/3)\oplus 2(3,1,1,4/3)\oplus (3,1,1,-2/3)\oplus 5(1,1,1,0)
\oplus 2(1,1,1,2)\oplus 2(1,1,1,-2)$,\\
where the chiral representations in $\psi(144)$ with respect to
R are those including the ordinary particles (without right-handed
neutrinos) and  the new exotic ones with labels
$(3,\bar{3},1,0)\oplus 2(1,\bar{3},1,-4/3)\oplus
(1,\bar{3},1,2/3)\oplus 2(3,1,1,4/3)\oplus (3,1,1,-2/3)$, all of them
belonging to the sector $\psi(1,1,6,\bar{6})$.

Normalizing the generators in G as stated before we have that
Eqs.(\ref{eq5}) and (\ref{eq6}) still hold with
$b_3=b_{3L}+b_{3R}$, where $b_{3L}$ and $b_{3R}$ are related to
SU(3)$_{CL}$ and SU(3)$_{CR}$ respectively. Then
$C_3(s)=C_{3R}(s)+C_{3L}(s)$.

R is broken down to \sem by $\phi_1(72)\oplus\phi_2(1296)$ with
the same VEVs as stated before. Now under the hypothesis that all the Higgs
fields in $\phi_1\oplus\phi_2$ contribute to the beta functions,
we have again that the different $C_i(s)$ contributions cancel out.
This time we get from
Eq.(\ref{eq5}) ln(M/M$_Z)=7.73\pm 0.15$ while from
Eq.(\ref{eq6}) we obtain ln(M/M$_Z)=
6.27\pm 0.31$, which are again incompatible.

On the other hand, the assumption that the extended
survival hypothesis\cite{esh} holds leads to
ln(M/M$_Z)=5.81\pm 0.11$ from  Eq.(\ref{eq5}) and to
 ln(M/M$_Z)=5.78\pm 0.28$  from Eq.(\ref{eq6})
which are consistent solutions.
The unification mass scale predicted now is M$\sim 3.3\times 10^4$ GeVs.
It is evident that this version of the model is rich in experimental
consequences.

The following list of comments refers to the model described by G and
$\psi(144)$ which breaks down to \sem with R as the only intermediate
gauge structure, properly implemented with the survival hypothesis\cite{sh},
the extended survival hypothesis\cite{esh}, and the modified horizontal
survival hypothesis\cite{ponce}:

\begin{itemize}
\item The evolution of the four gauge coupling constants in G meet
together in a single point at M$\sim 10^4$ GeVs, in good agreement
with precisions data test of the SM.
\item The two mass scales M$\sim 3.3\times 10^4$ GeVs and
M$_Z\sim 10^2$ GeVs are well within the reach of future experiments.
\item The only ordinary fermion which gets a tree level mass of order M$_Z$
is the t quark. It gets its mass via a flavor democratic mass matrix.
\item At the mass scale M$_Z$ the following exotic particles must exist:
8 {\it ``axigluons"}, two Up type quarks and one Down type quark.
\item The {\it queight} and the {\it quone} should get masses smaller
than M$_Z$.
\item The gauge fields not related to R and all the other exotic leptons
should get masses of order M$\sim 10^4$ GeVs.
\end{itemize}

\noindent
At first glance this version of the model could present the following
undesirable features:
\begin{itemize}
\item M$\sim 10^4$ GeVs could be a very small unification mass scale
(perhaps too close to the present limit for flavor changing neutral currents).
\item Since neither $\phi_1(72)$ or $\phi_2(1296)$ are able to produce a
tree level mass for the {\it queight} or the {\it quone}, those particles
can pick up only radiative or see-saw masses of a few GeVs (this
should be no problem if the {\it queight} is confined).
\item There is not a sufficient large mass scale in the model able to
generate see-saw mechanisms\cite{gellmann} for the three neutrinos
(most probably $\nu_e$ remains massless in this scheme as a
consequence of the symmetries of the vacuum as in the case of
the structure
[G$^V$, $\psi^V(108)$] discussed in Ref.\cite{ponce}).
\end{itemize}

The above mentioned three problems can be solved by the introduction of new
Higgs fields [for example $\phi_3(1,1,(\overline{15} +
\overline{21}), (15 + 21))]$
which give tree level masses of order M$_Z$ to the {\it queight}
and the {\it quone}. Then we can look for solutions
to the RGEs for the symmetry breaking chain

\begin{center}
G$\stackrel{M}{\longrightarrow}$R$\stackrel{M_{ch}}{\longrightarrow}$
\sm$\stackrel{M_Z}{\longrightarrow}$\sem,
\end{center}

\noindent
with the mass hierarchy M$>$M$_{ch}>$M$_Z$.
With two mass scales to be fixed and a lot of VEVs at our disposal
it is possible to look for solutions spanning the range
10$^7$ GeVs $\geq M>M_{ch}\geq 10^4 $ GeVs$ >M_Z \sim 10^2$ GeVs.

Now, independently of the existence of the unifying group G,
the set of fermion fields in $\psi(144)$ which is chiral with respect to R,
constitutes an anomaly-free chiral model with only three families,
different from the five models
(Marks I$-$V) introduced in Ref. \cite{gf}. Such a model deserves a
detailed study by its own sake.

\section{STABILITY OF THE PROTON}
In the subspace of the fundamental representation of
SU(6)$_{CR}\otimes$SU(6)$_{CL}$ the baryon number for G can be
associated with the $12\times 12$ diagonal matrix\\
B=$Dg.[(1/3,1/3,1/3,0,0,0)\oplus (1/3,1/3,1/3,0,0,0)]$. Since this
matrix does not correspond to a generator of G (neither of G$^\prime$),
then the baryon number is not gauged in the context of the models discussed
here.

Now due to the stated directions of the VEVs for $\phi_1$ and $\phi_2$,
it is clear that B$\langle\phi_1(72)\rangle$=
B$\langle\phi_2(1296)\rangle$=0. Therefore B is not broken
spontaneously by the
set of Higgs fields used for the breaking of R down to \sem. But what about
the set of scalars fields used for the breaking of G down to R? It can
be shown that it is possible to break G$\rightarrow$R using Higgs fields
$\phi_i=Z_4\phi_i(n_L,n_R,n_{CR},n_{CL})$ such that
B$\langle\phi_i(n_L,n_R,n_{CR},n_{CL})\rangle=0$, as long as
$n_K=1,6,\overline{6},15,\overline{15},21,\overline{21},35$ ($K=L,R,CR,CL)$
and as long as the directions for the VEVs of
SU(6)$_{CR}\otimes$SU(6)$_{CL}$ are such that\cite{langa}
$\alpha,\Delta\neq1,2,3$.
Examples of adequate Higgs fields and VEVs are presented for example in
Refs.\cite{ponce}. Our conclusion is that it is possible to choose
Higgs fields and VEVs which break G$\rightarrow$\sem, such that B is not
spontaneously broken.

To conclude that B is perturbatively conserved we follow
t'Hooft\cite{toof} and write B in the space of the fundamental representation
for SU(6)$_{CR}\otimes$SU(6)$_{CL}$ as B=(BL+$\Theta$)/2, where
BL=$Dg.[(1,1,1,-1,-1,-1)\otimes (1,1,1,-1,-1,-1)]$
is a generator of the G algebra which distinguishes baryon and
lepton number, and
$\Theta = Dg.[(1,1,1,1,1,1)\otimes(1,1,1,1,1,1)]$
generates a U(1)$_{\Theta}$ global symmetry
of the model. BL and $\Theta$ are both spontaneously
broken, but B is unbroken. A similar situation was analyzed in
Ref.\cite{ponce} for the structure [G$^V$, $\psi^V(108)$].

\section{CONCLUSIONS}
The model described by G$^\prime$ and $\psi^\prime(144)$ was studied originally
in Ref.\cite{elias} for the symmetry breaking pattern
G$\rightarrow$\sm$\rightarrow$\sem, under the assumption that all the fermion
fields (ordinary and mirrors) contribute to the RGEs and neglecting the
contribution of the Higgs fields. Substituting the experimental
values\cite{amal} of $\alpha_3(M_Z)$ and $\alpha_{EM}(M_Z)$ in the results
of Refs.\cite{elias} we find that the three gauge coupling constants
$\alpha_i, i=1,2,3$ of the SM do not meet at one point; i.e. those
results do not satisfy precision tests of the SM. Besides,
as we showed in the first section, it is impossible to implement the
survival hypothesis in this model due to the fact that
$\psi^\prime(144)$ is vectorlike with respect to G$^\prime$.

The new models we have studied here have the same gauge structure as the
model in Ref.\cite{elias}, but a different particle content. As a matter of
fact, $\psi(144)$ does not contain mirror fermion fields and it is not
vectorlike with respect to G. Therefore, the survival hypothesis can
be properly
implemented at each stage of the symmetry breaking pattern, and the
Appelquist-Carazzone\cite{apel} theorem can be properly used for the
decoupling of heavy fermion fields in the RGEs.

Numerical results were obtained here taking into account not
only the decoupling theorem and the survival hypothesis at each stage of the
breaking, but also including the effects of the scalar fields.
These effects
were calculated under two different assumptions and the results
were confronted with precision tests of the SM, with the
conclusion that under special circumstances the three gauge coupling
constants $\alpha_i ,i=1,2,3$ of the SM meet together at the
unification scale M without any intermediate mass scale above M$_Z$ [i.e.
without supersymmetry or extra physics beyond that contained in G and
$\psi(144)$].

The low unification scales discussed here
($10^7$ GeVs$ \geq$ M $\geq 10^5$ GeVs )do not conflict with data on proton
stability because baryon number is perturbatively conserved.
Also, lower energy
unification makes these models free from problems of grand unified
monopoles\cite{pbp} and the gauge hierarchy problem is also much less
severe (no fine tuning required?)

Finally let us see how the known mass spectrum for the elementary
fermion fields could be generated in the context of [G, $\psi(144)$]:

\begin{itemize}
\item The quark t acquire a tree level mass of order M$_Z$ by coupling
$\psi(\bar{6},1,1,6)\psi(1,6,\bar{6},1)$ to
$\phi_2(6,\bar{6},6,\bar{6})$ with the VEVs $\langle\phi_2\rangle$ as
stated in Section 2. The t quark (but not the b quark) gets
its mass via a flavor democratic mass matrix.
\item The b quark and $\tau$ lepton acquire see-saw masses of order
M$_Z^2/$M$_{ch}$ by coupling
$[\psi(\bar{6},1,1,6)+\psi(1,6,\bar{6},1)]\psi(1,1,6,\bar{6})$
to $\phi_1(6,1,\bar{6},1)+\phi(1,\bar{6},1,6)$ with the VEVs
$\langle\phi_1\rangle$ as stated in Section 2.
\item Masses for the charged fermion fields in the second and first
families can be generated as radiative corrections.
\end{itemize}

\noindent
These items are a novel realization of the horizontal survival
hypothesis\cite{hsh} according to which only the heaviest family
gets tree level masses from Yukawa couplings.
One aspect that the model does not clarify is the
observed smallness of the neutrino masses.

\section{ACKNOWLEDGMENTS}
This work was partially supported by CONACyT in M\'exico and Banco de la
Rep\'ublica in Colombia.

\pagebreak

\end{document}